\documentclass[
  aip,
  jcp,
onecolumn,
  superscriptaddress,
  10pt,
]{revtex4-2}

\usepackage{arydshln}
\usepackage{graphicx}
\usepackage{amsmath,amssymb}
\usepackage{bm}
\usepackage[colorlinks=true,linkcolor=blue,citecolor=blue,urlcolor=blue]{hyperref}
\usepackage{braket}
\usepackage{booktabs}
\usepackage{multirow}
\usepackage{ulem}
\usepackage[linesnumbered, ruled, longend]{algorithm2e}
\usepackage{multirow}
\usepackage{ulem}
\usepackage{lineno}
\usepackage[dvipsnames]{xcolor}
\usepackage{titlesec}

\renewcommand\thesubsubsection{\thesubsection\arabic{subsubsection}}
\titleformat{\subsubsection}
  {\normalfont\bfseries}  
  {\thesubsubsection.}
  {1em}
  {}

\begin{document}

\title{A Scalable Diagonalization Framework for Tensor-Product Bitstring Selected Configuration Interaction}

\author{Enhua Xu}
\email{enhua.xu@riken.jp}
\affiliation{RIKEN Center for Computational Science, Kobe, Japan}

\author{William Dawson}
\affiliation{RIKEN Center for Computational Science, Kobe, Japan}

\author{Himadri Pathak}
\affiliation{RIKEN Center for Interdisciplinary Theoretical and Mathematical Sciences, Wako, Japan}
\affiliation{RIKEN Center for Computational Science, Kobe, Japan}

\author{Takahito Nakajima}
\affiliation{RIKEN Center for Computational Science, Kobe, Japan}

\begin{abstract}
Selected configuration interaction (SCI) methods are effective for treating strongly correlated electronic systems, yet their scalability has long been limited by implementations that replicate the configuration interaction (CI) vector across processes, leading to severe memory bottlenecks.
Here, we present a fully distributed diagonalization framework tailored for extremely large selected determinant spaces, directly addressing this major scalability bottleneck of modern SCI methods.
The method is grounded in a tensor-product bitstring (TPB) representation, in which determinants are organized through a TPB structure constructed from selected $\alpha$- and $\beta$-bitstrings, and is referred to as tensor-product bitstring SCI (TBSCI).
An efficient TBSCI eigensolver is developed based on a novel bitstring-based Hamiltonian evaluation algorithm together with a suite of MPI communication strategies designed to improve parallel efficiency.
Large-scale full configuration interaction (FCI) benchmarks, employed as communication-intensive stress tests, demonstrate that the implemented TBSCI eigensolver continues to reduce the wall time for distributed diagonalization of 2.6 trillion determinants, reaching 54,000 nodes (more than 2.5 million cores) on supercomputer Fugaku.
Beyond scalability, we investigate the structural compactness of the TPB representation and show that selecting $\alpha$- and $\beta$-bitstrings according to their collective weights in a reference SCI wavefunction yields TPB-based wavefunctions approaching the FCI limit while using only a small fraction of determinants.
These results establish TBSCI as a scalable SCI methodology and provide evidence for the intrinsic compactness of the TPB representation.
\end{abstract}

\maketitle

\section{Introduction}

Accurately simulating strongly correlated quantum systems remains a central challenge in quantum chemistry. 
Full configuration interaction (FCI)~\cite{FCI_1984, FCI_1988, FCI_2000, FCI_2017} provides exact solutions to the electronic Schr\"odinger equation within a finite one-electron basis set.
The FCI wavefunction is formally expressed as
\begin{eqnarray}
\label{eq:FCI_C}
|\Psi_\mathrm{FCI}\rangle = \sum_{K=1}^{N_\mathrm{FCI}} c_K |D_K\rangle,
\end{eqnarray}
where each Slater determinant $|D_K\rangle$ is associated with a variational coefficient $c_K$, and the total number of determinants is 
$N_\mathrm{FCI} = \binom{M}{N_\alpha} \times \binom{M}{N_\beta}$, 
representing all possible distributions of $N_\alpha$ (spin-up) and $N_\beta$ (spin-down) electrons over $M$ spin-orbitals.
Enabled by advances in parallel algorithms and distributed data structures, modern FCI implementations have reached the trillion-determinant regime~\cite{FCI_2024, STPDAS_2024} and, more recently, extended to the quadrillion-determinant scale.~\cite{STPDAS_2025} 
Nevertheless, due to the factorial growth of $N_\mathrm{FCI}$ with system size, FCI is not intended as a general-purpose method for systems of broader chemical interest. 
This challenge has motivated the development of systematically improvable approximate methods.

Starting from Eq.~\eqref{eq:FCI_C}, one of the most direct approximations is to retain only the important determinants—those carrying the large weights ($c_K^2$) in the wavefunction—an idea that underlies selected configuration interaction (SCI) methods.
Over the past decades, several SCI methods have been proposed and developed, including configuration interaction using a perturbative selection made iteratively (CIPSI),~\cite{CIPSI} sparse FCI,~\cite{SFCI_2008} adaptive CI (ACI),~\cite{ACI_2016} adaptive sampling CI (ASCI),~\cite{ASCI_2016, ASCI_2023} semistochastic heat-bath CI (SHCI),~\cite{SHCI_2016, SHCI_2017, SHCI_2018} and iterative CI (iCI).~\cite{iCI_2016, iCI_2020, iCI_2021} 
Related ideas have also been explored in approaches based on a priori truncation,~\cite{deadwood_2001, deadwood_2009} as well as in more recent machine learning-based determinant selection strategies.~\cite{ML_2018, ML_2021, ML_2021_2, ML_2025}
These approaches have demonstrated that such selection strategies can yield highly compact wavefunctions for a wide range of molecular systems, enabling near-FCI accuracy with only a small fraction of the FCI determinants.

From an algorithmic perspective, a generic SCI calculation consists of three stages: (1) selecting important determinants, (2) diagonalizing the CI Hamiltonian within the selected subspace, and (3) applying perturbative corrections to recover residual correlation. 
Leveraging modern supercomputing architectures, several SCI methods~\cite{CIPSI, QUANTUM, ASCI_2016, ASCI_2023, SHCI_2016, SHCI_2017, SHCI_2018, iCI_2016, iCI_2020, iCI_2021} have implemented distributed-memory parallelization across all three stages, enabling determinant spaces to grow substantially beyond what was previously feasible. 
Yet in most existing implementations the CI vector remains fully replicated across nodes—not because the SCI framework inherently precludes distributed CI-vector storage, but because maintaining both efficiency and scalability with distributed CI-vector storage at large scale remains a nontrivial algorithmic and implementation challenge. As a consequence, reported SCI calculations typically involve up to $2 \times 10^{9}$ determinants in stage (2).~\cite{SHCI_2018}
As a variational method, SCI systematically improves the wavefunction description as more determinants are retained. 
For modern SCI calculations targeting increasingly accurate descriptions of large-scale strongly correlated systems, the distributed diagonalization of extremely large selected determinant subspaces therefore becomes the central scalability bottleneck.

This limitation motivates a reconsideration of how determinant spaces arising in SCI calculations may be structurally organized.

Consider that each determinant in Eq.~\eqref{eq:FCI_C} can be written as a tensor product of an $\alpha$- and a $\beta$-bitstring,
$|D_K\rangle = |S_w^\alpha\rangle \otimes |S_u^\beta\rangle$.
Accordingly, the FCI wavefunction can equivalently be expressed as
\begin{eqnarray}
\label{eq:FCI_TPB}
|\Psi_\mathrm{FCI}\rangle = \sum_{w=1}^{L_\alpha}\sum_{u=1}^{L_\beta} c_{(w,u)} 
|S_w^\alpha\rangle \otimes |S_u^\beta\rangle,
\end{eqnarray}
where $|S_{w}^{\alpha}\rangle$ and $|S_{u}^{\beta}\rangle$ denote the $\alpha$- and $\beta$-bitstrings, respectively, and $c_{(w,u)}$ is the associated variational amplitude, identical to the original FCI coefficient $c_K$ under a one-to-one mapping between $K$ and $(w,u)$; no factorization of amplitudes is implied.
$L_\alpha$ and $L_\beta$ are the numbers of all possible $\alpha$- and $\beta$-bitstrings, respectively, satisfying $N_\mathrm{FCI} = L_\alpha L_\beta$.
Hereafter, we refer to the representation in Eq.~\eqref{eq:FCI_TPB} as the tensor-product bitstring (TPB) representation.

The $\alpha/\beta$ factorization in the TPB representation has long been exploited in direct CI algorithms to enable efficient Hamiltonian application.~\cite{FCI_1984, FCI_1988, SFCI_2008} These approaches, however, typically assume the presence of complete sets of $\alpha$- and $\beta$-bitstrings.
In contrast, most SCI methods select individual determinants in a sparse and generally irregular manner, which breaks the tensor-product closure between the used $\alpha$- and $\beta$-bitstrings.
As a result, the underlying bitstring-level separability becomes difficult to exploit systematically in Hamiltonian evaluation, especially in conjunction with distributed CI-vector storage.

However, when determinants selected in SCI are embedded within the TPB representation, a structured bitstring-level indexing and connectivity pattern emerges.
We refer to this algorithmic organizational pattern as the TPB structure.
The TPB structure governs how determinants are indexed, traversed, and coupled in Hamiltonian application, independent of how determinants are selected within the TPB representation.
Within this organizing principle, fully distributed CI-vector storage enables efficient on-the-fly Hamiltonian evaluation in a scalable manner.
Building upon this structure, we formulate the tensor-product bitstring SCI (TBSCI) approach developed in this work.

Given the demonstrated compactness of determinant selection in SCI, a natural question then arises: \textit{does the TPB representation itself retain a comparable level of compactness in practice?}

In 2008, Surján {\it et al.}~\cite{SFCI_2008} selected important bitstrings by estimating their contributions to the energy and subsequently retained determinants in which at least one bitstring was deemed important. 
Their results indicated that the TPB representation can exhibit compactness under such a selection strategy for small molecular systems.
However, because their approach retains complete sets of $\alpha$- and $\beta$-bitstrings, its practical applicability becomes limited as $L_\alpha$ or $L_\beta$ grows large.
More recently, an IBM team selected important bitstrings based on their weights obtained by projecting quantum-computed wavefunctions, and then constructed determinants by taking all tensor products of the selected bitstrings.~\cite{IBM_2025} As an extension of the quantum selected configuration interaction (QSCI) strategy,~\cite{QSCI_2023} their primary objective in employing the TPB representation was to restore spin symmetry ($S^2$) in noisy quantum wavefunctions for closed-shell singlet states.
In both cases, the TPB representation was employed as a determinant construction scheme, while its structural implications for distributed CI-vector storage were not explored.

In this work, we adopt the same selection principle as in the IBM study, namely, extracting important $\alpha$- and $\beta$-bitstrings according to their weights in a reference wavefunction. Here, however, the reference wavefunction is generated on classical computers using a SCI method, thereby avoiding quantum noise while providing reproducible wavefunctions.
Once the important bitstrings are identified, we retain all determinants formed by taking the tensor products of the selected $\alpha$- and $\beta$-bitstrings, while excluding those that violate the adopted molecular point-group symmetry. 
This choice allows us to compare the coefficient distribution from all the symmetry-allowed determinants within the TPB representation with that of the FCI determinant space.
Based on these considerations, we develop a scalable TBSCI eigensolver compatible with fully distributed CI-vector storage. 
Its diagonalization framework is constructed upon the TPB structure, and through systematic refinement of its implementation and extensive large-scale testing, we demonstrate efficient parallel performance on supercomputer Fugaku across tens of thousands of compute nodes.
The implementation of perturbative corrections, corresponding to stage (3), is left for future work.

The remainder of this paper is organized as follows. In the Methodology section, we present the newly developed TPB-structured diagonalization framework in detail. In the Results section, we first demonstrate the scalability of the implemented TBSCI eigensolver on supercomputer Fugaku, and then describe the procedure for identifying important bitstrings, showing that the resulting TBSCI wavefunctions can approach the FCI limit using only a small fraction of the FCI determinants.
Finally, we conclude with a summary of the main findings and outline future prospects.

\section{Methodology}

We now present the diagonalization framework built upon the TPB structure.
For clarity of presentation, this subsection is organized into four parts, covering the distributed storage of the CI vector, the distributed matrix--vector multiplication, the efficient Hamiltonian computation algorithm tailored for TBSCI, and MPI communication optimization strategies.
These components are not independent modules but constitute a tightly integrated design, 
in which the chosen layout of the distributed CI vector serves as the foundation of both the distributed matrix--vector operations and the efficient Hamiltonian computation algorithm, 
and the design of the MPI communication optimization strategies is based on the data layout, computational workflow, and algorithmic formulation.

\subsection{Distributed storage of the CI vector}

Given the sets of important $\alpha$- and $\beta$-bitstrings, $\{S^{\alpha}\}=\{ S_{w}^{\alpha}\mid w=1,\dots,l_\alpha\}$ and $\{S^{\beta}\}=\{ S_{u}^{\beta}\mid u=1,\dots,l_\beta\}$, where $l_\alpha$ and $l_\beta$ denote the numbers of selected $\alpha$- and $\beta$-bitstrings, respectively, 
the complete tensor-product determinant set under the TPB representation is
\begin{eqnarray}
\label{eq:TPB_Det}
\mathcal{D}_\mathrm{TPB} =\left\{|S_{w}^{\alpha}\rangle\otimes|S_{u}^{\beta}\rangle \mid w=1,\dots,l_\alpha;\ u=1,\dots,l_\beta \right\},
\end{eqnarray}
which is referred to as the TPB determinant space.
Because each determinant is identified by its $(w,u)$ indices and the sets $\{S^{\alpha}\}$ and $\{S^{\beta}\}$ are stored on every process, determinants therefore need not be stored explicitly.
The CI vector is thus defined as the array of CI coefficients stored in the determinant index order.

The determinant indices are ordered such that the global index varies first with $\beta$ and then with $\alpha$.
Under this ordering, the CI vector decomposes naturally into segments, each associated with a fixed $\alpha$-bitstring.
To enable distributed storage, segments are assigned to MPI processes.
For a total of $N$ processes, process $p$ stores $m_p$ assigned $\alpha$-bitstrings (hence $m_p$ segments), with $\sum_{p=0}^{N-1} m_p = l_\alpha$.

\begin{figure}[h]
\centering
\includegraphics[width=0.9\textwidth]{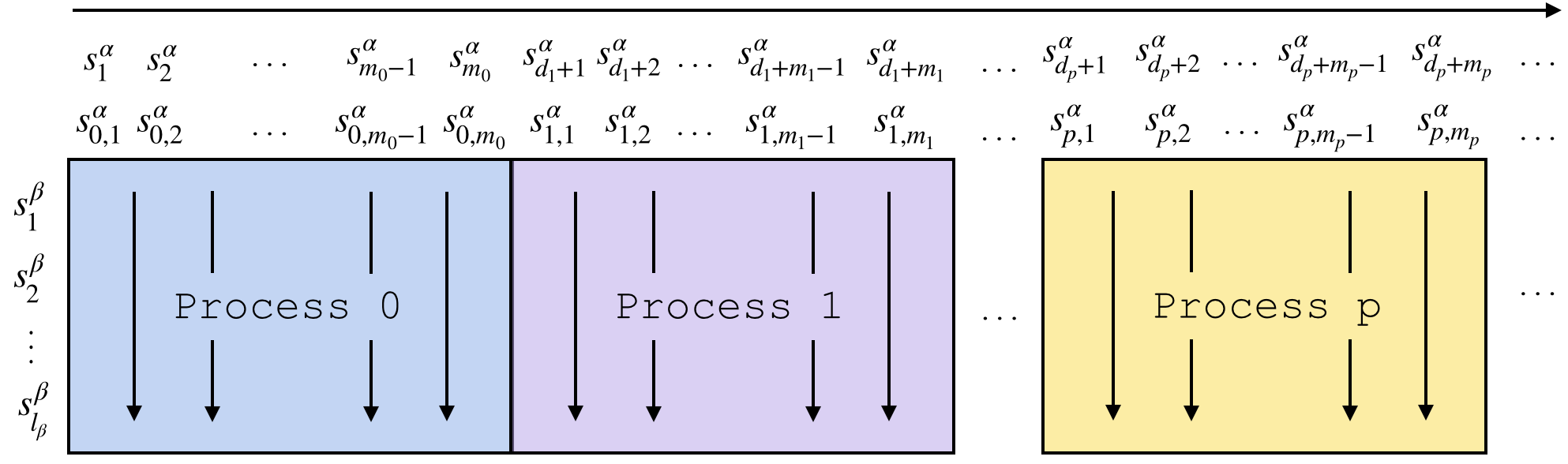}
\caption{Distribution of all determinants from $\mathcal{D}_\mathrm{TPB}$ into segments across $N$ processes (in this work each node runs exactly one process). Each process holds a strict subset of $m_p$ $\alpha$-bitstrings. The upper and lower rows of $\alpha$-bitstring labels indicate their global and local indices within each process, respectively. Here we define a displacement index $d_p = \sum_{i=0}^{p-1} m_i$ to indicate the offset of $\alpha$-bitstrings assigned to each process.
On a local process, the determinants are divided into segments (indicated by vertical arrows), where each segment contains all determinants sharing the same $\alpha$-bitstring.}
\label{fig:distribution}
\end{figure}

When all the determinants in $\mathcal{D}_\mathrm{TPB}$ are retained, each segment in the corresponding CI vector has the uniform length $l_\beta$.
In this case, the global index $K$ and the pair $(w,u)$ are related by
\begin{eqnarray}
w &=& \left\lfloor \frac{K-1}{l_\beta} \right\rfloor + 1, \quad
u = \mathrm{MOD}(K-1, l_\beta) + 1, \\
K &=& (w-1)l_\beta + u.
\end{eqnarray}
The corresponding segment-based distribution across processes is illustrated in Fig.~\ref{fig:distribution}.

In the general TBSCI setting, where determinant-level pruning is applied to $\mathcal{D}_\mathrm{TPB}$, the resulting determinant subset can be written as
\begin{eqnarray}
\mathcal{D}_\mathrm{TBSCI}=\left\{|S_w^\alpha\rangle\otimes|S_u^\beta\rangle
\mid w=1,\dots,l_\alpha;\ u\in\mathcal{I}_w^{(\beta)} \right\},
\end{eqnarray}
which is referred to as the TBSCI determinant space.
Here $\mathcal{I}_w^{(\beta)} \subseteq \{1,\dots,l_\beta\}$ denotes the index set of $\beta$-bitstrings paired with the $w$-th $\alpha$-bitstring. 
For a given $w$, $\mathcal{I}_w^{(\beta)}$ may range from the empty set to the full index set $\{1,\dots,l_\beta\}$.
Each segment therefore contains $|\mathcal{I}_w^{(\beta)}|$ determinants.
When $\mathcal{I}_w^{(\beta)} = \{1,\dots,l_\beta\}$ for all $w$, $\mathcal{D}_\mathrm{TBSCI} = \mathcal{D}_\mathrm{TPB}$.

Since segment lengths are no longer uniform, we store for each segment its global offset, owning process, and length $|\mathcal{I}_w^{(\beta)}|$.
In addition, for each process we store the global offset and total length of its local CI-vector portion.
All such segment-level structural metadata are replicated across processes, thereby providing a complete global view of the CI-vector layout.
In contrast, the associated $\beta$-indices in $\mathcal{I}_w^{(\beta)}$ are stored only on the owning process.
As a consequence, although the determinants are stored compactly in a determinant-by-determinant manner, each selected determinant, uniquely identified by its $(w,u)$ indices, can still be efficiently mapped to its position within the TPB determinant space, $\mathcal{D}_\mathrm{TPB}$.
This preservation of deterministic indexing enables direct and efficient addressing of determinants during Hamiltonian application under fully distributed CI-vector storage.
This design forms the structural basis of the TPB algorithmic framework.

In the present study, to investigate the compactness of the TPB representation, we exclude only the symmetry-forbidden determinants from $\mathcal{D}_\mathrm{TPB}$.
By pre-sorting the $\beta$-bitstrings according to their irreducible representations, the symmetry-allowed $\beta$-bitstrings form contiguous blocks within $\{S^\beta\}$, and each $\mathcal{I}_w^{(\beta)}$ corresponds to a contiguous interval.
Therefore, instead of storing the entire set $\mathcal{I}_w^{(\beta)}$, it suffices to store its starting index and length.

\subsection{Distributed matrix--vector multiplication}

TBSCI employs a Davidson-based diagonalization framework formulated under the TPB structure, in which the dominant computational task is the evaluation of the matrix--vector product \(\mathbf{W} = H \cdot \mathbf{U}\), where \(H\) denotes the Hamiltonian matrix and \(\mathbf{U}\) and \(\mathbf{W}\) represent CI vectors. When the CI vector is distributed such that each process \(p\) stores the local portion of the CI vector, denoted as \(\mathbf{U}_p\), the matrix--vector product naturally decomposes into local contributions. Each process computes its local contribution \(\mathbf{W}_p\) as
\begin{equation}
\label{eq:Wp=H*Uq}
\mathbf{W}_p = \sum_q H_{p,q} \cdot \mathbf{U}_q,
\end{equation}
where $H_{p,q}$ denotes the sub-block of $H$ that couples the rows owned by process $p$ with the columns corresponding to $\mathbf{U}_q$.
The efficient on-the-fly evaluation of Eq.~\eqref{eq:Wp=H*Uq} is described in the following subsection. Here, we focus on the data-access pattern and the computational workflow across distributed processes.

During each Davidson iteration, process \(p\), in principle, would need to fetch CI vector segments from all other processes. 
Each such fetch-and-compute operation (i.e., $\mathbf{W}_p \mathrel{+}= H \cdot \mathbf{U}_q$) is referred to as one \textit{step}, and thus a single Davidson iteration involves $N$ steps per process.
Throughout these \(N\) steps, processes operate asynchronously without global synchronization.
Since computation time varies across steps and processes, enforcing synchronization after each step would introduce unnecessary waiting.
Data transfers are performed via one-sided \texttt{MPI\_GET} operations, allowing each process to directly access remote memory without requiring active participation from the target process.

Moreover, to reduce the communication delays associated with \texttt{MPI\_GET} operations, computation and data transfer are overlapped.
On multi-core supercomputing architectures (e.g., Fugaku), one core on each node is dedicated to performing MPI data transfers for the next step, while the remaining cores execute the local computation of the current step using OpenMP.
Let \(t_{\mathrm{cal}}\) and \(t_{\mathrm{data}}\) denote the computation time and data retrieval time for a given step, respectively. If \(t_{\mathrm{data}} > t_{\mathrm{cal}}\), we define the delay time to the wall time as \(t_{\mathrm{delay}} = t_{\mathrm{data}} - t_{\mathrm{cal}}\). Otherwise, \(t_{\mathrm{delay}} = 0\), since MPI communication latency during data transfer—if it occurs—does not prolong the wall time. The total delay time \(T_{\mathrm{delay}}\) for one Davidson iteration is obtained by summing \(t_{\mathrm{delay}}\) over all \(N\) steps.

This analysis implies that efficient large-scale TBSCI diagonalization requires both fast Hamiltonian evaluation (small \(t_{\mathrm{cal}}\)) and low communication overhead (small \(t_{\mathrm{data}}\)). 
The following subsections address these requirements through a novel Hamiltonian evaluation algorithm together with a suite of communication strategies that enable scalable execution.

\subsection{Efficient Hamiltonian computation tailored for TBSCI}

Exploiting the TPB structure, we propose and implement an original algorithm that enables efficient on-the-fly Hamiltonian evaluation with memory-efficient data organization. The Hamiltonian matrix element between two determinants, $\langle D_{(w, u)} | \hat{H} | D_{(x, v)} \rangle$, can be written as $\langle S_{w}^\alpha | \langle S_{u}^\beta | \hat{H} | S_{x}^\alpha \rangle | S_{v}^\beta \rangle$. 
According to the Slater–Condon rules, this matrix element is nonzero only if
\begin{eqnarray}
\mathrm{DIFF}(S_w^\alpha, S_x^\alpha) + \mathrm{DIFF}(S_u^\beta, S_v^\beta) \leq 2,
\end{eqnarray}
where $\mathrm{DIFF}(S, S')$ denotes the number of differing spin-orbitals between bitstrings $S$ and $S'$.
We classify the allowed cases into six categories $[a,b]$, where $a = \mathrm{DIFF}(S_w^\alpha, S_x^\alpha)$ and $b = \mathrm{DIFF}(S_u^\beta, S_v^\beta)$. 
The six combinations $[2,0]$, $[1,1]$, $[1,0]$, $[0,2]$, $[0,1]$, and $[0,0]$ cover all nonzero contributions. The first five combinations correspond to off-diagonal elements evaluated on-the-fly, while the last combination ($[0,0]$) represents diagonal elements, which are also computable on-the-fly but are precomputed and stored to improve efficiency.

Although only a subset of determinants in the TPB determinant space $\mathcal{D}_\mathrm{TPB}$ is retained, TBSCI preserves the tensor-product indexing structure associated with the TPB representation. 
Importantly, the TPB structure serves as an organizing principle for indexing and Hamiltonian traversal; it does not require the retained determinants to form a tensor-product closure of $\mathcal{D}_\mathrm{TPB}$.
As a consequence, for a fixed pair $(S_w^\alpha,S_x^\alpha)$, all $\beta$-side excitation candidates can be generated within the selected $\beta$-bitstring set $\{S^\beta\}$ and then filtered by the corresponding $\beta$-index sets $\mathcal{I}_w^{(\beta)}$ and $\mathcal{I}_x^{(\beta)}$.
This property allows excitation relations among $\beta$-bitstrings to be precomputed once and reused across all $\alpha$-bitstring pairs under the TPB indexing structure.
Specifically, we construct the \texttt{BETA\_SINGLE\_LINK} and \texttt{BETA\_DOUBLE\_LINK} link tables (implemented as arrays), which store single and double excitation connectivity within $\{S^\beta\}$.
As a result, for each admissible $(S_w^\alpha,S_x^\alpha)$ pair, the allowed Hamiltonian contributions are evaluated by traversing excitation links within $\{S^\beta\}$ rather than enumerating determinant pairs in the full product space.

Building on this structure, Algorithm~\ref{alg:interaction_terms} computes $\mathbf{W}_p \mathrel{+}= H_{p,q} \cdot \mathbf{U}_q$.
For each pair $(S_w^\alpha,S_x^\alpha)$, the value of $a=\mathrm{DIFF}(S_w^\alpha,S_x^\alpha)$ is computed (Line 3), after which the allowed excitation contributions involving $\{S^\beta\}$ are generated directly from the precomputed \texttt{BETA\_SINGLE\_LINK} and \texttt{BETA\_DOUBLE\_LINK} link tables.
Excitation traversal produces candidate $\beta$-bitstrings independently of segment boundaries and therefore requires efficient membership testing in $\mathcal{I}_w^{(\beta)}$.
In general, efficient membership testing for $S_u^\beta \in \mathcal{I}_w^{(\beta)}$ can be achieved using hash-based structures or sorted index lists with binary search.
In the present implementation, since all symmetry-allowed determinants in $\mathcal{D}_\mathrm{TPB}$ are retained, each $\mathcal{I}_w^{(\beta)}$ forms a contiguous interval in $\{S^\beta\}$, so membership testing reduces to a simple interval check on irreducible representations.

\begin{algorithm}[!ht]
\DontPrintSemicolon
\SetAlgoNoLine
\SetArgSty{textup}
\KwData{Current process $p$, current step $q$; BETA\_SINGLE\_LINK and BETA\_DOUBLE\_LINK lists}
\For{$S_w^\alpha \leftarrow S^{\alpha}_{p,1}$ \KwTo $S^{\alpha}_{p,m_p}$}{
  \For{$S_x^\alpha \leftarrow S^{\alpha}_{q,1}$ \KwTo $S^{\alpha}_{q,m_q}$}{
    case $\gets$ \textbf{DIFF}($S_w^\alpha$, $S_x^\alpha$)\;
    \uIf{case $=2$}{
      \ForEach{$S_v^\beta \in \mathcal{I}_x^{(\beta)} $}{
        $S_u^\beta \gets S_v^\beta$\;
        If $S_u^\beta \in \mathcal{I}_w^{(\beta)}$, calculate $\langle S_{w}^\alpha | \langle S_{u}^\beta | \hat{H} | S_{x}^\alpha \rangle | S_{v}^\beta \rangle$ as [2,0] term\;
      }
    }
    \uElseIf{case $=1$}{
      \ForEach{$S_v^\beta \in \mathcal{I}_x^{(\beta)} $}{
        \ForEach{$S_u^\beta \in \text{BETA\_SINGLE\_LINK}(:,S_v^\beta)$}{
          If $S_u^\beta \in \mathcal{I}_w^{(\beta)}$, calculate $\langle S_{w}^\alpha | \langle S_{u}^\beta | \hat{H} | S_{x}^\alpha \rangle | S_{v}^\beta \rangle$ as [1,1] term\;
        }
        $S_u^\beta \gets S_v^\beta$\;
        If $S_u^\beta \in \mathcal{I}_w^{(\beta)}$, calculate $\langle S_{w}^\alpha | \langle S_{u}^\beta | \hat{H} | S_{x}^\alpha \rangle | S_{v}^\beta \rangle$ as [1,0] term\;
      }
    }
    \ElseIf{case $=0$}{
      \ForEach{$S_v^\beta \in \mathcal{I}_x^{(\beta)} $}{
        \ForEach{$S_u^\beta \in \text{BETA\_SINGLE\_LINK}(:,S_v^\beta)$}{
          If $S_u^\beta \in \mathcal{I}_w^{(\beta)}$, calculate $\langle S_{w}^\alpha | \langle S_{u}^\beta | \hat{H} | S_{x}^\alpha \rangle | S_{v}^\beta \rangle$ as [0,1] term\;
        }
        $S_u^\beta \gets S_v^\beta$\;
        If $S_u^\beta \in \mathcal{I}_w^{(\beta)}$, calculate $\langle S_{w}^\alpha | \langle S_{u}^\beta | \hat{H} | S_{x}^\alpha \rangle | S_{v}^\beta \rangle$ as [0,0] term\;
      }
    }
  }
  \BlankLine
  $S_x^\alpha \leftarrow S_w^\alpha$\;
  \ForEach{$S_v^\beta \in$ the $q$-th batch of $\{S^\beta\}$}{
    \ForEach{$S_u^\beta \in \text{BETA\_DOUBLE\_LINK}(:,S_v^\beta)$}{
      If $S_u^\beta \in \mathcal{I}_w^{(\beta)}$, calculate $\langle S_{w}^\alpha | \langle S_{u}^\beta | \hat{H} | S_{x}^\alpha \rangle | S_{v}^\beta \rangle$ as [0,2] term\;
    }
  }
}
\caption{Local computation of matrix elements on process $p$ at step $q$.}
\label{alg:interaction_terms}
\end{algorithm}

The effective sparsity of single and double excitations can be approximately estimated as $\sqrt{N_{\mathrm{TBSCI}}/N_{\mathrm{FCI}}}$, where $N_{\mathrm{TBSCI}}$ is the number of determinants retained in TBSCI.
Accordingly, the memory usage of \texttt{BETA\_SINGLE\_LINK} and \texttt{BETA\_DOUBLE\_LINK} can be estimated to scale as $l_\beta N_{\mathrm{occ}} N_{\mathrm{vir}} \sqrt{N_{\mathrm{TBSCI}}/N_{\mathrm{FCI}}}$ and $l_\beta N_{\mathrm{occ}}^2 N_{\mathrm{vir}}^2 \sqrt{N_{\mathrm{TBSCI}}/N_{\mathrm{FCI}}}$, respectively, for \(N_{\mathrm{occ}}\) occupied and \(N_{\mathrm{vir}}\) virtual orbitals.
The \texttt{BETA\_DOUBLE\_LINK} array can be very large and is therefore partitioned into batches, one batch per process, with each batch containing a subset of $\{S^\beta\}$.
For each $S_u^\beta$, its column \texttt{BETA\_DOUBLE\_LINK(:, $S_u^\beta$)} is precomputed and stored entirely on a single process with auxiliary arrays recording the corresponding index ranges to enable fast access.
The computation of the [0,2] term is thus carried out batch by batch during one Davidson iteration (Lines 27–32 of Algorithm~\ref{alg:interaction_terms}).
Since this [0,2] term involves only a loop over $S_w^\alpha$ and does not reference $S_x^\alpha$, each process fetches the relevant \texttt{BETA\_DOUBLE\_LINK} batch only once per step, removing memory pressure while introducing negligible additional communication volume.

In Algorithm~\ref{alg:interaction_terms}, the sparsity factor ($\sqrt{N_{\mathrm{TBSCI}} / N_{\mathrm{FCI}}}$) appears once in the computations of the $[2,0]$ and $[0,2]$ terms, but twice in that of the $[1,1]$ term. 
Consequently, the dominant contributions arise from the $[2,0]$ and $[0,2]$ terms, each scaling as 
$N_{\mathrm{TBSCI}} \cdot N_{\mathrm{occ}}^2 \cdot N_{\mathrm{vir}}^2 \cdot \sqrt{N_{\mathrm{TBSCI}} / N_{\mathrm{FCI}}}$, whereas the $[1,1]$ term scales as $N_{\mathrm{TBSCI}} \cdot N_{\mathrm{occ}}^2 \cdot N_{\mathrm{vir}}^2 \cdot (N_{\mathrm{TBSCI}} / N_{\mathrm{FCI}})$.
In this work, the TBSCI program is also employed to perform FCI calculations, where the sparsity factor becomes exactly one and the computational dependence simplifies to $N_{\mathrm{FCI}} \cdot N_{\mathrm{occ}}^2 \cdot N_{\mathrm{vir}}^2$, with dominant contributions arising from the $[2,0]$, $[1,1]$, and $[0,2]$ terms.
A summary of the scaling behavior and the dominant terms is provided in Table~\ref{tab:scaling}.
It should be noted that the traditional Handy–Knowles determinant-based FCI algorithm scales as $N_{\mathrm{FCI}} \cdot N_{\mathrm{occ}} \cdot N_{\mathrm{vir}}$,~\cite{FCI_1984, FCI_1988} which is more favorable for complete active-space calculations where the full determinant space is retained. 
More recently, the small-tensor-product distributed active space (STP-DAS) framework~\cite{STPDAS_2025} has further extended FCI calculations to the quadrillion-determinant regime by introducing distributed active-space decompositions together with lossless compression of CI vectors, while maintaining high numerical accuracy.
The present work is not intended to compete with existing FCI methodologies,~\cite{FCI_1984, FCI_1988, FCI_2000, FCI_2017, FCI_2024, STPDAS_2024, STPDAS_2025} but to employ the FCI workload as a communication-intensive stress test for the distributed TBSCI eigensolver. 
As already reflected in Algorithm~\ref{alg:interaction_terms}, the method is designed to operate on arbitrary determinant subsets of $\mathcal{D}_\mathrm{TPB}$. 
Its objective is therefore fundamentally different: rather than accelerating FCI, it enables scalable diagonalization of extremely large selected determinant spaces under fully distributed CI-vector storage.

\begin{table}[htbp]
\centering
\caption{Computational scaling and dominant terms of SCI and FCI calculations in the framework of TBSCI.}
\label{tab:scaling}
\begin{ruledtabular}
\begin{tabular}{lcc}
Method & Overall scaling & Dominant terms \\
\colrule
SCI & 
$N_{\mathrm{SCI}} N_{\mathrm{occ}}^{2} N_{\mathrm{vir}}^{2} 
\sqrt{N_{\mathrm{SCI}} / N_{\mathrm{FCI}}}$ & 
$[2,0]$, $[0,2]$ \\
FCI & 
$N_{\mathrm{FCI}} N_{\mathrm{occ}}^{2} N_{\mathrm{vir}}^{2}$ & 
$[2,0]$, $[1,1]$, $[0,2]$ \\
\end{tabular}
\end{ruledtabular}
\end{table}

In existing SCI implementations that evaluate Hamiltonian elements on-the-fly, the mixed-spin $[1,1]$ term typically dominates the computational cost.
For a fixed $\alpha$-pair with $\mathrm{DIFF}(S_w^\alpha,S_x^\alpha)=1$, evaluating the $[1,1]$ term requires looping over all pairs $(S_u^\beta,S_v^\beta)$ with $u \in \mathcal{I}_w^{(\beta)}$ and $v \in \mathcal{I}_x^{(\beta)}$, which leads to a quadratic traversal over the corresponding $\beta$-bitstring index sets.
In contrast, by traversing precomputed excitation connectivity within the selected $\beta$-bitstring set $\{S^\beta\}$, the dominant computational cost grows approximately in proportion to the number of retained determinants (as shown in Table~\ref{tab:scaling}), because determinant pairs are no longer enumerated explicitly and each retained determinant participates in only a limited number of excitation connections determined by the Slater–Condon excitation structure.
We note that when $|\mathcal{I}_w^{(\beta)}|$ is small relative to the typical excitation connectivity size, the constant-factor overhead associated with connectivity traversal and membership testing may exceed that of direct pair enumeration.
In the present implementation, however, each $\mathcal{I}_w^{(\beta)}$ forms a contiguous interval, so membership testing reduces to a simple interval check and the above constant-factor consideration does not arise.

\subsection{MPI Communication Optimization Strategies}

As described above, the distributed matrix–vector multiplication involves frequent inter-node data transfers. 
Although the underlying SCI formulation is not inherently tied to a specific communication strategy, communication contention can dominate the execution time at large node counts, resulting in degraded MPI efficiency and, in extreme cases, even an increase in wall time with increasing node count. 
To maintain parallel efficiency at large node counts, we developed and integrated a suite of MPI communication optimization strategies within the distributed TBSCI eigensolver.

\subsubsection{Avoiding unnecessary MPI data transfers.}
As described in Eq.~\eqref{eq:Wp=H*Uq}, each process \( p \) must retrieve CI-vector segments from other processes during distributed matrix--vector multiplication. However, not all remote segments are required for local Hamiltonian evaluation.
Specifically, for any remote \(\alpha\)-bitstring \( S_x^\alpha \), if all local \(\alpha\)-bitstrings \(\{S_{p,1}^\alpha, \dots, S_{p,m_p}^\alpha\}\) satisfy \(\mathrm{DIFF}(S_{p,i}^\alpha, S_x^\alpha) > 2\), then the segment corresponding to \( S_x^\alpha \) can be safely omitted. 
This excitation-connectivity information is precomputed and stored, allowing each process to bypass unnecessary data transfers and thereby reduce MPI communication volume to scale approximately with the local computational workload.

\subsubsection{Exploiting molecular symmetry.}  
In addition to excitation-based pruning, molecular symmetry provides a further mechanism for eliminating unnecessary communication.
Specifically, if $\mathrm{DIFF}(S_w^\alpha, S_x^\alpha)=2$ and $S_w^\alpha$ and $S_x^\alpha$ belong to different irreducible representations, the corresponding Hamiltonian matrix elements vanish identically, and the associated CI vector segments do not need to be fetched.
Exploiting D$_{2h}$ symmetry, for instance, reduces communication and computational costs by a factor of $\sim 64$.

\subsubsection{Minimizing long-distance MPI communication.}  
Each node is assigned a unique ID from \(0\) to \(N-1\). Since messages between nodes with nearby IDs traverse fewer network hops (when suitably ordered), we sort \(\alpha\)-bitstrings by their excitation levels relative to the Hartree--Fock (HF) $\alpha$-bitstring and assign them to nodes in ascending order. Lower-excitation bitstrings are placed on lower-numbered nodes, while higher-excitation ones are assigned to higher-numbered nodes.
Because excitation differences between distant bitstrings often exceed two, data retrieval from remote nodes is frequently unnecessary. This excitation-aware mapping ensures that, even on tens of thousands of nodes, each node primarily fetches CI vector segments from its nearest neighbors, thereby mitigating MPI congestion and sustaining high parallel efficiency.

\subsubsection{Balancing memory and computational load.}  
Algorithm~\ref{alg:interaction_terms} allows estimation of computational costs associated with each $\alpha$-bitstring segment. The total workload of a process is approximated by summing the costs of its assigned $\alpha$-bitstrings.  
While \(\alpha\)-bitstrings are assigned in ascending excitation order, balancing memory usage and computational load simultaneously is challenging. We define two expansion factors: the memory expansion factor (the ratio of assigned bitstring count to the average) and the computation expansion factor (the ratio of estimated cost to the average).
Strictly enforcing memory balance (memory expansion factor $\approx 1$) risks severe computational imbalance, causing idle processes, whereas strictly enforcing computational balance (computation expansion factor $\approx 1$) may overload memory on some nodes.  
We therefore adopt a compromise strategy, maintaining both factors moderately above one to balance memory and computation effectively.  
For FCI calculations, where per-segment costs are uniform, a uniform distribution naturally balances both aspects.

\subsubsection{Absorbing delays via reassignment of $[0,2]$ tasks.}  
Since data retrieval and computation are executed concurrently, delays ($t_{\mathrm{delay}} > 0$) are more likely to occur in steps where the computational time $t_{\mathrm{cal}}$ is relatively small. 
The $[0,2]$ terms (double excitations within the $\beta$-bitstring space) exhibit a favorable pattern in this regard: they incur substantial computational cost but require negligible data retrieval, and are evaluated independently in a separate batch loop (lines 27--32 in Algorithm~\ref{alg:interaction_terms}).
We therefore treat the $[0,2]$ tasks as a computational reservoir. Each process estimates the computational workload of its $N$ steps based on the assigned $\alpha$-bitstrings and reassigns its $N$ batches of $[0,2]$ tasks to steps with smaller pre-estimated workloads, thereby absorbing some potential delays across Davidson steps.
Steps that require neither remote fetches of CI-vector segments nor local $[0,2]$ computations (after reassignment) can be eliminated entirely.
While this breaks the strict one-step-one-batch structure in Algorithm~\ref{alg:interaction_terms}, the original form is maintained for clarity.

\subsubsection{Odd-even process fetch ordering.}  
Simultaneous MPI data transfers can cause network congestion spikes. 
To alleviate this issue, the target ranks are sorted according to the number of required segments. 
Odd-numbered processes fetch data in ascending order, while even-numbered processes fetch in descending order.
This simple ordering effectively reduces both the frequency and magnitude of delay times, especially in large-scale runs on tens of thousands of nodes.

\subsubsection{Check-if-busy dynamic scheduling.}  
When process A performs \texttt{MPI\_GET} from process B, both are marked \texttt{BUSY}. Other processes preferentially fetch from \texttt{NOT BUSY} targets but can access \texttt{BUSY} ones if necessary.  
This dynamic scheduling reduces contention and improves overall parallel efficiency.

\subsubsection{Sleep strategy under severe MPI congestion.}  
Despite prior optimizations, severe $T_{\mathrm{delay}}$ spikes occasionally occur in runs involving more than five thousand nodes, leading to cascading congestion akin to traffic jams.  
To mitigate this, if a process detects $t_{\mathrm{delay}}$ exceeding a specified threshold (e.g., 0.2~s), it enters a brief sleep period (e.g., 0.1~s), marking itself as \texttt{BUSY} during the sleep to discourage other processes from fetching data from itself.
While this approach slightly increases $T_{\mathrm{delay}}$ within a single Davidson iteration, it consistently suppresses large delay spikes and improves performance stability in large-scale runs.  
Threshold values were empirically tuned on Fugaku and may require adjustment for other architectures or computational methods.

\section{Results}

\subsection{Scalability and Parallel Efficiency}

To assess the parallel performance of the TBSCI eigensolver under the most communication-intensive conditions, we deliberately benchmarked the code using FCI calculations. 
Within the same TPB-based diagonalization framework, FCI calculations—corresponding to the limiting case of SCI with a sparsity factor of unity—provide a particularly stringent stress test for both the distributed CI-vector architecture and the associated MPI communication strategies.

Under this setting, we computed FCI energies for N$_2$ (aug-cc-pVDZ) and CN (cc-pVTZ) potential energy surfaces (PESs), as well as single-point energies of Cr$_2$ (STO-3G) and N$_2$ (cc-pVTZ) near equilibrium. 
Their FCI dimensions under the frozen-core approximation and molecular symmetry are summarized in Table~\ref{tab:systems}. These large-scale calculations also provide useful benchmark data for future developments in quantum chemistry.
The HF reference states and corresponding FCIDUMP files were generated using PySCF~\cite{PYSCF}, and all data—including the CI vector and intermediate arrays—were kept entirely in memory throughout the calculations. 
This in-memory design avoids disk I/O overhead during iterative updates.

\begin{table}[htbp]
\centering
\caption{System information used in TBSCI benchmark calculations.}
\label{tab:systems}
\begin{ruledtabular}
\begin{tabular}{lcccc}
System  & Number of frozen orbitals & Active space & Molecular symmetry & FCI dimension \\
\colrule
N$_2$ (aug-cc-pVDZ)   & 2  & CAS(10e, 44o) & D$_{2h}$ & $1.47 \times 10^{11}$ \\
CN (cc-pVTZ)          & 2  & CAS(9e, 58o)  & C$_{2v}$& $4.86 \times 10^{11}$ \\
Cr$_2$ (STO-3G)      & 12 & CAS(24e, 24o) & D$_{2h}$ & $9.14 \times 10^{11}$ \\
N$_2$ (cc-pVTZ)       & 2  & CAS(10e, 58o) & D$_{2h}$ & $2.62 \times 10^{12}$ \\
\end{tabular}
\end{ruledtabular}
\end{table}

\begin{figure}[!ht]
\centering
\includegraphics[width=1\columnwidth]{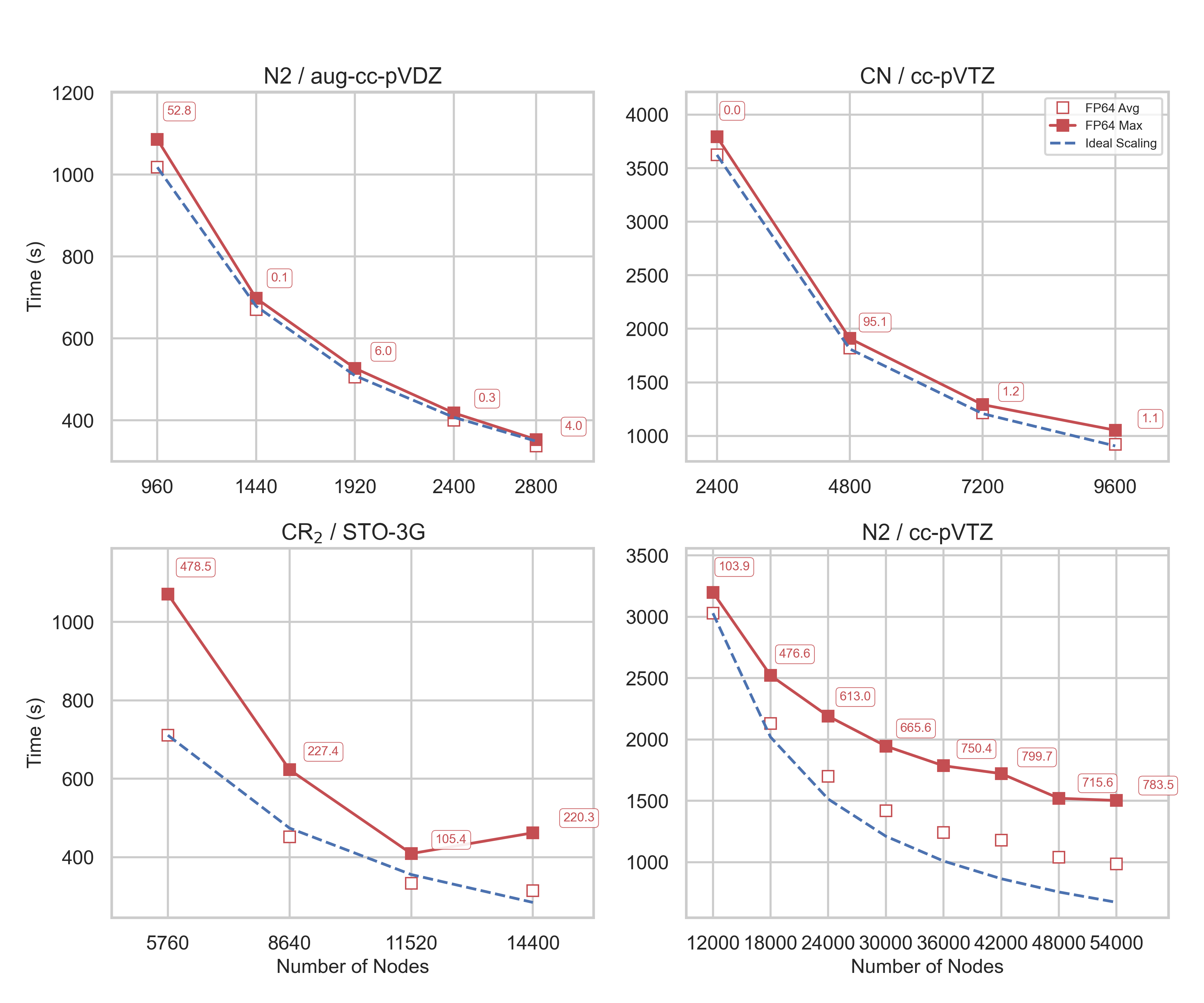}
\caption{Wall time (in seconds) for a single distributed matrix--vector multiplication across four benchmark systems. 
Open squares and filled squares denote the average and maximum wall times across nodes, respectively. 
Small boxed labels indicate $T_{\text{delay}}$ on the slowest process (in seconds). 
We plot the ideal scaling curve based on the initial average wall-time data point to quantify the overhead of our implementation.}
\label{fig:timing}
\end{figure}

Figure~\ref{fig:timing} presents the wall time of a single distributed matrix--vector multiplication—the dominant cost in Davidson diagonalization—across increasing node counts for all systems. Detailed timings and node-hour costs are provided in Table~S1 in the \textbf{Supplementary Material}. Due to the substantial computational cost of these large-scale runs, the timing metrics are reported from the initial production run rather than averaged over multiple trials.
For N$_2$ (aug-cc-pVDZ) and CN (cc-pVTZ), the maximum and average wall times remain very close across all node counts, closely following the ideal strong-scaling trend. This indicates that communication delays remain minor and that the distributed eigensolver operates in a computation-dominated regime.
For Cr$_2$ (STO-3G), communication delays become increasingly significant at higher node counts. 
Although the maximum wall time decreases when increasing the node count from 5760 to 11520, a further increase to 14400 nodes results in a noticeable rise in the maximum wall time, indicating that communication contention begins to limit further strong-scaling gains for this system.
The largest test is N$_2$ (cc-pVTZ), involving more than $2.6 \times 10^{12}$ determinants. Although the gap between the maximum and average wall times generally increases with node count, the overall wall time continues to decrease up to 54,000 nodes (more than 2.5 million cores on Fugaku). 
Even at the largest scale tested, communication overhead increases substantially, yet the iteration cost remains primarily determined by computational work.
Interestingly, despite being the largest system in terms of determinant count, N$_2$ (cc-pVTZ) does not exhibit the most severe relative communication delays, suggesting that runtime communication behavior depends not only on system size but also on detailed excitation structure and workload distribution.

All reported timings correspond to a single Davidson iteration. 
In practice, once the diagonalization is launched, the per-iteration wall time remains nearly constant throughout the iterative procedure. 
Consequently, the total runtime can be reliably estimated from the per-iteration cost and the number of Davidson iterations required for convergence, indicating stable runtime characteristics of the distributed TBSCI implementation at large node counts.
In the present work we employ the standard Davidson algorithm as described in Ref.~\onlinecite{QUANTUM}, without additional acceleration techniques. 
As a representative example, along the PES of N$_2$ (aug-cc-pVDZ), the number of Davidson iterations required for convergence increases significantly as the bond is stretched. 
At $R = 2.068$~a.u., convergence is achieved in 12 iterations, with a per-iteration wall time of about 698~s on 1,440 nodes, corresponding to a total wall time of approximately 2 hours. 
At $R = 4.2$~a.u., where convergence is slower, 52 Davidson iterations are required, with a per-iteration wall time of about 420~s on 2,400 nodes, giving a total wall time of approximately 6 hours.

Furthermore, we examined empirical cost scaling at fixed computational resources by performing calculations for four representative molecular systems spanning a wide range of estimated Hamiltonian computation costs, $N_{\mathrm{FCI}} N_{\mathrm{occ}}^2 N_{\mathrm{vir}}^2$. 
All calculations were carried out using 1000 compute nodes on Fugaku. 
The effective computation time per Davidson iteration, defined as the wall time excluding the communication delay $T_{\text{delay}}$, increases approximately in proportion to the estimated Hamiltonian computation cost, confirming the effectiveness of the algorithmic design.
Detailed numerical data are provided in Table~S2 of the \textbf{Supplementary Material}.

These results demonstrate that the TBSCI eigensolver sustains scalable distributed diagonalization under extreme concurrency, with computation remaining dominant over communication up to tens of thousands of nodes.

Finally, we note that mixed-precision strategies have been explored in quantum chemistry and large-scale eigenvalue computations.~\cite{Pakiari1975, Carson2018, Pokhilko2018, Tsai2022} In this context, a possible variant of TBSCI would perform the matrix--vector multiplication in single precision while retaining double precision for all other operations. Such a design could reduce communication volume and facilitate future GPU-oriented implementations. 
A systematic investigation of its numerical accuracy and parallel performance in the TBSCI framework will be presented in future work.

\subsection{Compactness of the TPB representation from SCI-derived bitstrings}

Having validated the scalability of the distributed diagonalization framework, we now examine whether the TPB representation admits compact approximations to correlated wavefunctions in practice.

We employ the widely used \texttt{DICE} package~\cite{SHCI_2016, SHCI_2017} to generate the SCI reference wavefunction.
In the heat-bath CI procedure, candidate determinants whose estimated coefficients exceed a predefined threshold $\theta$ are selected for diagonalization;~\cite{SHCI_2016} here $\theta$ is fixed at $10^{-5}$.
After normalizing the resulting SCI eigenvector, the weight of an $\alpha$- or $\beta$-bitstring is defined as the sum of squared coefficients over all determinants containing that bitstring.
The $\alpha$- and $\beta$-bitstrings are ranked by these weights, and important bitstrings are selected using a relative threshold $\delta$ referenced to the corresponding HF bitstring weights.
Varying $\delta$ yields different selected bitstring sets.
For each selected set, we perform a variational TBSCI calculation by including all symmetry-allowed determinants in $\mathcal{D}_\mathrm{TPB}$.

Table~\ref{tab:N2_PES} reports TBSCI results for N$_2$ (aug-cc-pVDZ) along the potential energy surface (PES) using SCI-derived bitstrings selected by relative weight thresholds $\delta = 10^{-6}$, $10^{-7}$, $10^{-8}$, $10^{-9}$, and $0$.
Despite the transition of the electronic structure from predominantly single-reference near equilibrium to strongly multi-reference in the dissociation regime, TBSCI yields energies that systematically and smoothly approach the FCI limit as $\delta$ is reduced.
At $\delta = 10^{-8}$, the deviations $E_{\mathrm{TBSCI}} - E_{\mathrm{FCI}}$ are already below $2$~m$E_h$, and sub-millihartree accuracy is reached at $\delta = 10^{-9}$ while using well below $1\%$ of the FCI determinants.
Across the entire PES, the deviations decrease monotonically with $\delta$ and exhibit only weak dependence on bond length, indicating a stable and well-controlled convergence pattern in this test case.
Overall, these results indicate that bitstring-weight screening induces a systematic refinement path under the TPB representation, consistent with its compactness in these test cases.

\begin{table}[!htbp]
\centering
\caption{Summary of TBSCI results for N$_2$ (aug-cc-pVDZ) along the PES using SCI-derived bitstrings. 
All energy differences $E_{\mathrm{TBSCI}} - E_{\mathrm{FCI}}$ are in millihartree (mH). 
Here, $\delta$ denotes the relative bitstring-weight threshold: an $\alpha$- or $\beta$-bitstring is selected if its weight exceeds $\delta$ times the corresponding HF bitstring weight.}
\label{tab:N2_PES}
\begin{ruledtabular}
\begin{tabular}{cccccc}
$\delta$  & $N_{\mathrm{bitstrings}}$ & $N_{\mathrm{TBSCI}}$ & $N_{\mathrm{TBSCI}}/N_{\mathrm{FCI}}$ & $E_{\mathrm{TBSCI}} - E_{\mathrm{FCI}}$ (mH) & (nodes, seconds) \\
\colrule
\multicolumn{6}{c}{$R = 2.068$ a.u. \hspace{0.6cm} $E_{\mathrm{HF}}=-108.961045$ \hspace{0.6cm} $E_{\mathrm{FCI}}=-109.295263$} \\
\colrule
$1\times10^{-6}$ & 1,523     & 572,235       & 0.00039\%   & 14.591 & (12, 0.47) \\
$1\times10^{-7}$ & 5,480     & 4,810,418     & 0.0033\%    & 5.296 & (12, 5.0)\\
$1\times10^{-8}$ & 15,776    & 38,387,174    & 0.026\%     & 1.616 & (12, 49.3)\\
$1\times10^{-9}$ & 38,620    & 228,641,788   & 0.16\%      & 0.475 & (12, 370.1)\\
0                & 124,480   & 2,872,138,682 & 1.95\%      & 0.088 & (36, 2111.1)\\
\colrule
\multicolumn{6}{c}{$R = 2.4$ a.u. \hspace{0.6cm} $E_{\mathrm{HF}}=-108.875323$ \hspace{0.6cm} $E_{\mathrm{FCI}}=-109.258928$} \\
\colrule
$1\times10^{-6}$ & 1,870   & 742,908       & 0.00050\% & 15.249 & (12, 0.70) \\
$1\times10^{-7}$ & 6,375   & 6,480,851     & 0.0044\%  & 5.386 & (12, 8.1) \\
$1\times10^{-8}$ & 17,853  & 48,556,961    & 0.033\%   & 1.682 & (12, 75.0) \\
$1\times10^{-9}$ & 42,647  & 272,638,925   & 0.18\%    & 0.512 & (12, 545.7)\\
0                & 142,390 & 3,354,713,882 & 2.28\%    & 0.081 & (36, 3197.5)\\
\colrule
\multicolumn{6}{c}{$R = 2.7$ a.u. \hspace{0.6cm} $E_{\mathrm{HF}}=-108.748038$ \hspace{0.6cm} $E_{\mathrm{FCI}}=-109.182160$} \\
\colrule
$1\times10^{-6}$ & 2,240   & 1,008,784     & 0.00068\%  & 15.725 & (12, 1.2)\\
$1\times10^{-7}$ & 7,441   & 8,705,921     & 0.0059\%   & 5.534 & (12, 14.1)\\
$1\times10^{-8}$ & 20,242  & 61,107,582    & 0.041\%    & 1.739 & (12, 122.4)\\
$1\times10^{-9}$ & 47,918  & 336,845,184   & 0.23\%     & 0.525 & (12, 891.2)\\
0                & 157,027 & 3,842,317,677 & 2.61\%     & 0.080 & (36, 4993.0)\\
\colrule
\multicolumn{6}{c}{$R = 3.0$ a.u. \hspace{0.6cm} $E_{\mathrm{HF}}=-108.618929$ \hspace{0.6cm} $E_{\mathrm{FCI}}=-109.108790$} \\
\colrule
$1\times10^{-6}$ & 2,748   & 1,444,334     & 0.00098\%  & 16.239 & (12, 2.3)\\
$1\times10^{-7}$ & 9,105   & 12,703,807    & 0.0086\%   & 5.520 & (12, 25.9)\\
$1\times10^{-8}$ & 24,013  & 83,584,179    & 0.057\%    & 1.737 & (12, 226.2)\\
$1\times10^{-9}$ & 55,410  & 437,097,908   & 0.30\%     & 0.531 & (12, 1568.1)\\
0                & 173,593 & 4,454,632,169 & 3.02\%     & 0.081 & (36, 8005.7)\\
\colrule
\multicolumn{6}{c}{$R = 3.6$ a.u. \hspace{0.6cm} $E_{\mathrm{HF}}=-108.401280$ \hspace{0.6cm} $E_{\mathrm{FCI}}=-109.016333$} \\
\colrule
$1\times10^{-6}$ & 4,780   & 3,618,030     & 0.0025\%  & 14.391 & (12, 11.1)\\
$1\times10^{-7}$ & 14,058  & 27,870,742    & 0.019\%   & 4.996 & (12, 117.2)\\
$1\times10^{-8}$ & 36,022  & 175,911,036   & 0.12\%    & 1.503 & (12, 994.6)\\
$1\times10^{-9}$ & 79,476  & 840,877,816   & 0.57\%    & 0.437 & (12, 3153.6)\\
0                & 206,186 & 5,761,751,654 & 3.91\%    & 0.078 & (48, 15261.6)\\
\colrule
\multicolumn{6}{c}{$R = 4.2$ a.u. \hspace{0.6cm} $E_{\mathrm{HF}}=-108.242916$ \hspace{0.6cm} $E_{\mathrm{FCI}}=-108.986039$} \\
\colrule
$1\times10^{-6}$ & 6,912   & 6,855,544     & 0.0047\%  & 11.440 & (12, 35.7)\\
$1\times10^{-7}$ & 21,399  & 61,044,875    & 0.041\%   & 3.502 & (12, 426.5)\\
$1\times10^{-8}$ & 50,284  & 333,094,790   & 0.23\%    & 1.034 & (12, 3095.3)\\
$1\times10^{-9}$ & 106,006 & 1,450,439,300 & 0.98\%    & 0.289 & (24, 9237.9)\\
0                & 220,324 & 6,324,862,976 & 4.29\%    & 0.071 & (48, 26496.6)\\
\end{tabular}
\end{ruledtabular}
\end{table}

The same bitstring-selection procedure was further applied to CN, Cr$_2$, and N$_2$ (cc-pVTZ), with the results summarized in Table~\ref{tab:OTHERS_TBSCI}. 
For CN, several geometries along the PES were examined, while single-point calculations were performed for Cr$_2$ and N$_2$ (cc-pVTZ).
Across all systems, TBSCI systematically approaches the FCI limit as the bitstring selection threshold $\delta$ is reduced.
For the molecular systems considered here, sub-millihartree accuracy is achieved at $\delta = 10^{-10}$ while using well below $0.56\%$ of the FCI determinants. 
Although the required fraction $N_{\mathrm{TBSCI}}/N_{\mathrm{FCI}}$ is larger for the strongly correlated transition-metal system Cr$_2$ than for main-group molecules, near-FCI accuracy is still recovered with a substantially reduced determinant space. 
Reference FCI energies for CN with cc-pVDZ, cc-pVTZ, and complete basis set (CBS) extrapolation along the PES are listed in Table~S3 of the \textbf{Supplementary Material}.

\begin{table}[!htbp]
\centering
\caption{Summary of TBSCI results for CN (cc-pVTZ), Cr$_2$ (STO-3G), and N$_2$ (cc-pVTZ) using SCI-derived bitstrings. 
For CN, an open-shell system, the numbers of selected $\alpha$- and $\beta$-bitstrings ($l_\alpha$ and $l_\beta$) are not identical and are therefore listed separately as $l_\alpha/l_\beta$. 
All energy differences $E_{\mathrm{TBSCI}} - E_{\mathrm{FCI}}$ are in millihartree (mH). 
The bond length ($R$) is in \AA, except for N$_2$, which has a bond length of 2.068 a.u.\ ($\approx$1.094~\AA) to match the potential energy surface in Table~\ref{tab:N2_PES}. 
Here, $\delta$ denotes the relative bitstring-weight threshold: an $\alpha$- or $\beta$-bitstring is selected if its weight exceeds $\delta$ times the corresponding Hartree--Fock bitstring weight.}
\label{tab:OTHERS_TBSCI}
\begin{ruledtabular}
\begin{tabular}{cccccc}
$\delta$ & $N_{\mathrm{bitstrings}}$ & $N_{\mathrm{TBSCI}}$ & 
$N_{\mathrm{TBSCI}}/N_{\mathrm{FCI}}$ & $E_{\mathrm{TBSCI}} - E_{\mathrm{FCI}}$ (mH) & (nodes, seconds) \\
\colrule
\multicolumn{6}{c}{CN \hspace{0.5cm} $R = 1.116$~\AA \hspace{0.5cm} $E_{\mathrm{HF}}=-92.219490$ \hspace{0.5cm} $E_{\mathrm{FCI}}=-92.569252$}  \\
\colrule
$1\times10^{-6}$ & 1,984/1,652     & 1,415,808      & 0.00029\% & 17.985 & (12, 2.6) \\
$1\times10^{-7}$ & 6,630/5,040     & 11,207,206     & 0.0023\%  & 6.510 & (12, 34.0) \\
$1\times10^{-8}$ & 19,914/13,362   & 79,673,340     & 0.016\%   & 2.151 & (12, 390.6)\\
$1\times10^{-9}$ & 52,187/30,663   & 469,503,223    & 0.097\%   & 0.677 & (12, 3106.0)\\
0                & 305,847/125,366 & 12,654,398,243 & 2.60\%    & 0.065 & (144, 14019.6)\\
\colrule
\multicolumn{6}{c}{CN \hspace{0.5cm} $R = 1.5$~\AA \hspace{0.5cm} $E_{\mathrm{HF}}=-92.059594$ \hspace{0.5cm} $E_{\mathrm{FCI}}=-92.476056$} \\
\colrule
$1\times10^{-6}$ & 2,477/1,919     & 1,768,335      & 0.00036\%  & 17.844 & (12, 4.3) \\
$1\times10^{-7}$ & 8,279/5,621     & 15,071,035     & 0.0031\%   & 6.181 & (12, 59.6) \\
$1\times10^{-8}$ & 23,647/14,219   & 98,778,650     & 0.020\%    & 2.030 & (12, 584.8)\\
$1\times10^{-9}$ & 59,523/30,862   & 525,680,797    & 0.11\%     & 0.645 & (12, 4177.4)\\
0                & 311,092/114,033 & 10,549,477,483 & 2.17\%     & 0.066 & (144, 13893.2)\\
\colrule
\multicolumn{6}{c}{CN \hspace{0.5cm} $R = 2.0$~\AA \hspace{0.5cm} $E_{\mathrm{HF}}=-91.844595$ \hspace{0.5cm} $E_{\mathrm{FCI}}=-92.324826$} \\
\colrule
$1\times10^{-6}$ & 5,292/2,965     & 4,743,784      & 0.00098\%   & 13.904 & (12, 37.8)\\
$1\times10^{-7}$ & 15,933/8,039    & 36,786,548     & 0.0076\%    & 4.788 & (12, 346.8)\\
$1\times10^{-8}$ & 42,851/18,919   & 218,222,058    & 0.045\%     & 1.470 & (12, 3739.1)\\
$1\times10^{-9}$ & 97,277/37,180   & 956,531,522    & 0.20\%      & 0.456 & (24, 13296.6)\\
0                & 357,906/109,454 & 10,318,500,738 & 2.12\%      & 0.060 & (144, 43692.4)\\
\colrule
\multicolumn{6}{c}{Cr$_2$ \hspace{0.5cm} $R = 1.5$~\AA \hspace{0.5cm} $E_{\mathrm{HF}}=-2064.078900$ \hspace{0.5cm} $E_{\mathrm{FCI}}=-2064.808195$} \\
\colrule
$1\times10^{-6}$ & 5,293   & 3,820,595      & 0.00042\%  & 24.535 & (12, 6.0)\\
$1\times10^{-7}$ & 17,326  & 40,118,388     & 0.0044\%   & 9.346 & (12, 98.1)\\
$1\times10^{-8}$ & 47,239  & 291,894,399    & 0.032\%    & 3.425 & (12, 1287.5)\\
$1\times10^{-9}$ & 107,597 & 1,495,787,547  & 0.16\%     & 1.417 & (24, 9897.9)\\
$1\times10^{-10}$& 199,571 & 5,116,950,531  & 0.56\%     & 0.803 & (36, 27027.4)\\
0	             & 328,795 & 13,906,467,547 & 1.52\%     & 0.582 & (144, 27230.9)\\

\colrule
\multicolumn{6}{c}{N$_2$ \hspace{0.5cm} $R = 2.068$~a.u. \hspace{0.5cm} $E_{\mathrm{HF}}=-108.984093$ \hspace{0.5cm} $E_{\mathrm{FCI}}=-109.375153$} \\
\colrule
$1\times10^{-6}$ & 1,530   & 586,634        & 0.000022\% & 20.668 & (12, 0.60)\\
$1\times10^{-7}$ & 5,472   & 5,147,844      & 0.00020\%  & 8.036 & (12, 6.0)\\
$1\times10^{-8}$ & 16,861  & 42,751,597     & 0.0016\%   & 2.633 & (12, 71.8)\\
$1\times10^{-9}$ & 43,914  & 288,849,332    & 0.011\%    & 0.844 & (12, 623.5)\\
0                & 260,795 & 12,554,326,295 & 0.48\%     & 0.087 & (144, 3772.6)\\

\end{tabular}
\end{ruledtabular}
\end{table}

\begin{figure}[!ht]
\centering
\includegraphics[width=1\columnwidth]{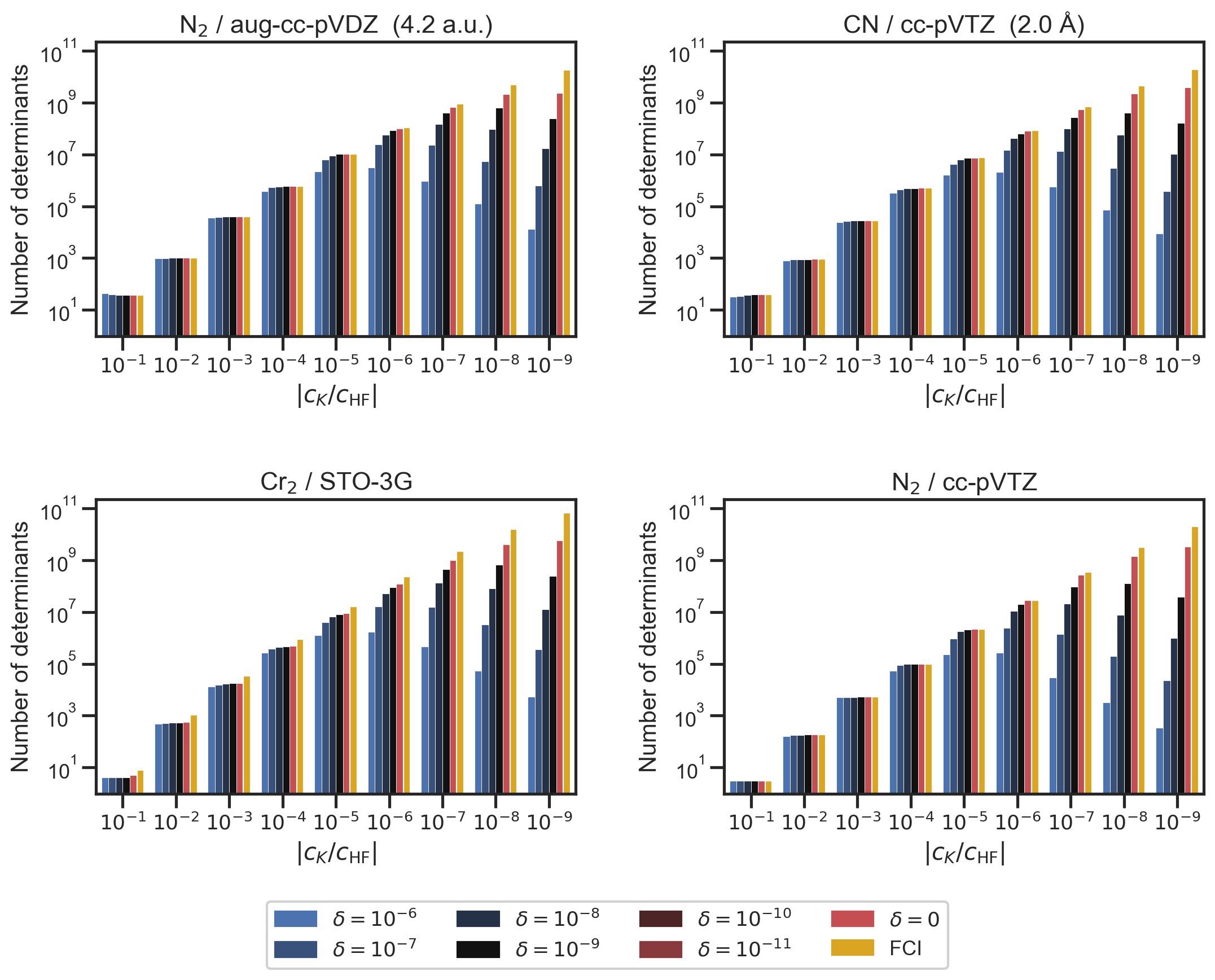}
\caption{Distributions of CI coefficient ratios $|c_K/c_{\mathrm{HF}}|$ within the TBSCI determinant spaces constructed from SCI-derived bitstrings.
Bars report the number of determinants in each logarithmic bin.
Different colors correspond to relative bitstring-weight thresholds $\delta$ used to select $\alpha$- and $\beta$-bitstrings,
while the reference bars denote the FCI distribution under identical symmetry and frozen-core conditions.}
\label{fig:CI_histograms}
\end{figure}

To further characterize how the TBSCI determinant space $\mathcal{D}_\mathrm{TBSCI}$ evolves as the bitstring-weight threshold $\delta$ is tightened, we analyze the distribution of CI coefficients in the resulting TBSCI wavefunctions.
For each system and each $\delta$, we construct $\mathcal{D}_\mathrm{TPB}$ from the SCI-derived $\alpha$- and $\beta$-bitstrings, retain all symmetry-allowed determinants to define the corresponding $\mathcal{D}_\mathrm{TBSCI}$, solve the resulting TBSCI eigenproblem, and bin the absolute coefficient ratios $|c_K/c_{\mathrm{HF}}|$ on a logarithmic scale.
Within each bin, we count the number of determinants whose coefficients fall into that range.
The same analysis is performed for the corresponding FCI eigenvector, providing a direct reference distribution.
Figure~\ref{fig:CI_histograms} illustrates a consistent trend across the tested systems, with one representative geometry shown for each system.
Additional coefficient-distribution plots for the remaining geometries are provided in Figure~S1 of the \textbf{Supplementary Material}.
Even at relatively loose bitstring screening (large $\delta$), the corresponding $\mathcal{D}_\mathrm{TBSCI}$ already contains most of the determinants with the largest coefficients in the FCI eigenvector.
As $\delta$ is reduced, additional determinants primarily populate progressively smaller-coefficient bins, and the overall histogram approaches the FCI reference distribution.
This behavior suggests that, for these systems, determinants with the largest FCI coefficients exhibit a structured organization within a determinant space induced by the TPB representation from a small number of important  $\alpha$- and $\beta$-bitstrings.
We emphasize that the present study is based on a limited set of molecular systems; assessing the breadth and limitations of this structural compactness across broader chemical problems remains an important direction for future work.

Taken together, Tables~\ref{tab:N2_PES} and~\ref{tab:OTHERS_TBSCI} and the coefficient-distribution analysis in Fig.~\ref{fig:CI_histograms} show that selecting $\alpha$- and $\beta$-bitstrings by their collective weights in a SCI reference wavefunction yields systematically improvable TBSCI determinant spaces, whose variational energies approach the FCI limit while retaining only a small fraction of the determinants.
Because the present study retains all symmetry-allowed tensor products of the selected bitstrings, the resulting TBSCI wavefunctions are not expected to be as compact as determinant-selected SCI wavefunctions (e.g., those produced by \texttt{DICE}), which explicitly exclude many low-contributing determinants.
Here our goal is to assess the compactness of the TPB representation itself under a bitstring-weight selection principle, rather than to maximize determinant-space compactness through additional determinant-level screening.
Such determinant-level refinement within the TPB representation is left for future work.

To further examine TBSCI energies for systems beyond the practical reach of FCI, we consider larger basis sets for Cr$_2$ (Ahlrichs SV)~\cite{SV} and N$_2$ (cc-pVQZ), for which benchmark energies are taken from high-accuracy DMRG (density matrix renormalization group)~\cite{DMRG_2015} and FCIQMC (full configuration interaction quantum Monte Carlo)~\cite{i-FCIQMC} calculations, respectively.
The results are summarized in Table~\ref{tab:BIG_TBSCI}.
In contrast to the smaller-basis cases, reducing the bitstring threshold to $\delta = 10^{-9}$ does not yet yield sub-millihartree accuracy.
Only at $\delta = 0$, by retaining all bitstrings present in the reference SCI wavefunction, do the TBSCI energies approach the benchmark values to within approximately 1~m$E_h$.
In the N$_2$ (cc-pVQZ) case, the best TBSCI energy is slightly lower than the reported FCIQMC value. 
However, this observation does not imply superior overall performance of TBSCI relative to FCIQMC. 
For larger and more strongly correlated systems, a purely variational TBSCI treatment without perturbative correction is expected to be less competitive than FCIQMC.
For completeness, the corresponding TBSCI coefficient distributions (without FCI reference distributions) are also provided in Figure~S1 of the \textbf{Supplementary Material}.

\begin{table}[!htbp]
\centering
\caption{Summary of TBSCI results for Cr$_2$ (Ahlrichs SV) and N$_2$ (cc-pVQZ) using SCI-derived bitstrings. 
All energy differences $E_{\mathrm{TBSCI}} - E_{\mathrm{benchmark}}$ are in millihartree (mH). 
Here, $\delta$ denotes the relative bitstring-weight threshold: an $\alpha$- or $\beta$-bitstring is selected if its weight exceeds $\delta$ times the corresponding Hartree--Fock bitstring weight.}
\label{tab:BIG_TBSCI}
\begin{ruledtabular}
\begin{tabular}{cccccc}
$\delta$ & $N_{\mathrm{bitstrings}}$ & $N_{\mathrm{TBSCI}}$  & $E_{\mathrm{TBSCI}} - E_{\mathrm{benchmark}}$ (mH) & (nodes, seconds) \\
\colrule
\multicolumn{6}{c}{Cr$_2$ \hspace{0.5cm} $R = 1.5$~\AA \hspace{0.5cm} $E_{\mathrm{HF}}=-2086.572971$ \hspace{0.5cm} $E_{\mathrm{DMRG}}=-2086.444784$} \\
\colrule
$1\times10^{-6}$ & 5,669     & 5,994,031           & 41.233  & (12, 11.2)\\
$1\times10^{-7}$ & 20,756    & 71,784,176          & 19.042 & (12, 180.9)\\
$1\times10^{-8}$ & 65,895    & 679,551,775         & 8.372 & (12, 2409.7)\\
$1\times10^{-9}$ & 181,976   & 5,061,656,122       & 3.756 & (48, 6770.8)\\
$1\times10^{-10}$& 427,692   & 27,568,299,482      & 1.871 & (386, 7228.9)\\
0	             & 1,722,583 & 490,574,659,035     & 0.919 & (7200, 10279.3)\\

\colrule
\multicolumn{6}{c}{N$_2$ \hspace{0.5cm} $R = 2.068$~a.u. \hspace{0.5cm} $E_{\mathrm{HF}}=-108.991735$ \hspace{0.5cm} $E_{\mathrm{FCIQMC}}=-109.40521$} \\
\colrule
$1\times10^{-6}$ & 1,916   & 887,486           & 30.961 & (12, 1.28)\\
$1\times10^{-7}$ & 7,268   & 10,169,032        & 13.662 & (12, 20.5)\\
$1\times10^{-8}$ & 24,819  & 100,565,129        & 4.798 & (12, 306.5)\\
$1\times10^{-9}$ & 73,249  & 815,730,549        & 1.423 & (12, 3554.9)\\
0                & 826,347 & 146,434,093,579    & -0.310& (2880, 10311.1) \\
\end{tabular}
\end{ruledtabular}
\end{table}

In practical SCI calculations, the parallel performance characteristics differ from the FCI limiting case.
Because only a subset of bitstrings is selected, the connectivity density of the Hamiltonian decreases, reducing inter-process communication intensity. At the same time, segment-level computational costs become nonuniform, which can lead to significant imbalance in wall times across processes. As a result, load imbalance can become more limiting than the communication delay $T_{\mathrm{delay}}$.
For example, in the Cr$_2$ (Ahlrichs SV) calculation using 7200 nodes, a purely memory-balanced distribution yields an average per-iteration wall time of 161~s across processes, whereas the maximum wall time reaches 734~s. After applying the overall balancing strategy, the average wall time remains 161~s, but the maximum wall time is reduced to 312~s. This comparison shows that the overall balancing strategy substantially mitigates load imbalance.
Nevertheless, the remaining gap between the average and maximum wall times indicates that segment workloads are intrinsically difficult to predict accurately. For example, the computational costs for evaluating the $[2,0]$ and $[0,2]$ terms are similar in principle, but in practice their effective workloads can differ substantially due to differences in data-access patterns and link traversal. Further improvements in load balancing beyond the current pre-estimation-based strategy will be investigated in future work.

In addition, practical TBSCI calculations typically exhibit a rapid growth of $N_{\mathrm{TBSCI}}$ as $\delta$ is reduced, which can quickly increase the cost of the variational diagonalization.
For Cr$_2$ (Ahlrichs SV), for example, tightening $\delta$ from $10^{-6}$ to $0$ increases $N_{\mathrm{TBSCI}}$ from $\sim 6\times 10^{6}$ to $\sim 4.9\times 10^{11}$ determinants.
Correspondingly, the runtime rises from 11.2~s (12 nodes; 30 iterations) to 2.9~h (7200 nodes; 37 iterations) in the present implementation.
This behavior highlights the practical motivation for incorporating perturbative corrections: the leading determinants should be treated variationally, while the residual correlation from the much larger set of secondary determinants is more economically recovered by a non-iterative second-order correction.

\subsection{Memory footprint of replicated link tables}

The memory footprint of TBSCI can be broadly grouped into four components.
(i) The one- and two-electron integrals, which are system dependent.
In the present study their sizes remain moderate and are stored locally; for substantially larger systems, a distributed storage strategy may become necessary.
(ii) The CI vectors used in Davidson diagonalization.
In our implementation, a restart scheme requires storing at most $2N_{\mathrm{max}}+3$ vectors (with $N_{\mathrm{max}}$ the maximum subspace dimension before restart).
Since these vectors are fully distributed, their memory cost per node is not a limiting factor.
(iii) MPI environment memory, which consumes relatively little memory and is difficult to estimate a priori.
(iv) TPB-based link data structures that enable efficient on-the-fly Hamiltonian evaluation under fully distributed CI-vector storage.

Among these, \texttt{BETA\_DOUBLE\_LINK} is stored in a distributed manner, whereas \texttt{BETA\_SINGLE\_LINK} is replicated across processes.
In the present implementation, each entry of the \texttt{BETA\_SINGLE\_LINK} table stores the index of a $\beta$-bitstring $S_u^{\beta}$ singly connected to $S_v^{\beta}$ (4-byte integer), together with a sign factor (stored as a 1-byte integer) and two orbital-difference indices. 
Although the orbital indices would in general require 2-byte integers, the number of active orbitals considered here is sufficiently small that they can be safely stored as 1-byte integers.
This results in a per-entry storage cost of 7 bytes in the current implementation.
Within the TPB-based on-the-fly Hamiltonian evaluation framework, replicated storage of the \texttt{BETA\_SINGLE\_LINK} table is structurally required to maintain computational efficiency under fully distributed CI-vector storage. Eliminating this replication would lead to substantial recomputation overhead in the inner loop.

Table~\ref{tab:SINGLE_MEM} reports the memory footprint of \texttt{BETA\_SINGLE\_LINK} for the largest calculation of each system listed in Tables~\ref{tab:N2_PES}, \ref{tab:OTHERS_TBSCI}, and \ref{tab:BIG_TBSCI}.
For the FCI benchmark workloads, the replicated \texttt{BETA\_SINGLE\_LINK} table becomes sizable and reaches 7.92~GB for N$_2$ (cc-pVTZ).
In contrast, for practical SCI calculations (e.g., Cr$_2$ with Ahlrichs SV and N$_2$ with cc-pVQZ), the reduced Hamiltonian connectivity leads to substantially smaller \texttt{BETA\_SINGLE\_LINK} tables.
In typical open-shell systems, the number of distinct $\alpha$-bitstrings tends to exceed that of $\beta$-bitstrings in the selected set; constructing the link table on the $\beta$ side therefore minimizes the replicated memory footprint.

To further illustrate the practical memory usage, we consider the N$_2$ (cc-pVTZ) FCI benchmark using 30,000 nodes.
The runtime system report indicates that the maximum memory usage per node is approximately 23~GB.
For this calculation, the distributed CI vectors (with $N_{\max}=8$) account for approximately 12.4~GB per node, while the replicated \texttt{BETA\_SINGLE\_LINK} table contributes roughly 8~GB.
This example demonstrates that diagonalizations involving extremely large determinant spaces can still be carried out within practical memory limits on modern supercomputers.

\begin{table}[htbp]
\centering
\caption{Number of selected $\beta$-bitstrings ($l_\beta$) and memory footprint of \texttt{BETA\_SINGLE\_LINK} in the largest calculation performed for each system. 
The upper four rows correspond to FCI benchmark workloads (sparsity factor equal to unity), while the lower rows correspond to SCI calculations with reduced connectivity.}
\label{tab:SINGLE_MEM}
\begin{ruledtabular}
\begin{tabular}{cccc}
System & Basis set & $l_{\beta}$  & Memory (GB) \\
\colrule
N$_2$ & (aug-cc-pVDZ)   & 211,771,560  & 1.38  \\
CN    & (cc-pVTZ)       &  91,642,320  & 0.60  \\
Cr$_2$ & (STO-3G)       & 389,398,464  & 2.54  \\
N$_2$ & (cc-pVTZ)      &1,214,260,740  & 7.92  \\
\midrule
Cr$_2$ & (Ahlrichs SV)  & 119,402,360 &  0.78  \\
N$_2$ & (cc-pVQZ)      &  115,211,618  & 0.75  \\
\end{tabular}
\end{ruledtabular}
\end{table}

\section{Conclusions and Prospects}

In this work, we have developed a distributed TBSCI eigensolver built upon the TPB representation and its associated structural organization of determinants.
By combining distributed CI-vector storage, excitation-aware Hamiltonian evaluation, and MPI communication optimization strategies, the implementation sustains stable distributed diagonalization of up to $2.6 \times 10^{12}$ determinants on 54,000 nodes of supercomputer Fugaku.
Beyond scalability, we have examined the structural properties of the TPB representation and shown that, when bitstrings are selected according to their collective weights in a reference SCI wavefunction, the resulting TBSCI wavefunctions closely approach the FCI limit using only a small fraction of the determinants.
Taken together, these results establish TBSCI as an SCI framework grounded in the TPB representation, in which compactness is a property of the representation while scalability is enabled by the TPB structure.

Looking forward, the next stage of development concerns refining determinant selection within the TPB representation to further compress the resulting TBSCI wavefunction.
One possible route is to apply FCIQMC-style stochastic sampling~\cite{FCIQMC, i-FCIQMC} within the TPB determinant space, using walker populations as a practical proxy for determinant importance.
In the present work, all calculations were performed using canonical Hartree--Fock orbitals. Since natural orbitals~\cite{NO_1955, NO_1972} and split-localized orbitals~\cite{split_2003} have been shown to provide more compact CI expansions, incorporating such optimized orbitals into the TPB representation is expected to further enhance its compactness and will be explored in future work.
Another important direction concerns the incorporation of second-order perturbative corrections (PT2).
In principle, PT2 may be formulated on top of the TBSCI wavefunction.
As a non-iterative procedure, PT2 does not require explicit storage of the full interacting determinant space, and the contributing determinants can be generated and evaluated in batches.
Achieving this while maintaining high parallel efficiency and mitigating communication overhead will require further algorithmic developments and program optimization, potentially including semi-stochastic strategies.~\cite{SHCI_2017, SHCI_2018}
Finally, further improvements in parallel efficiency will be pursued, for example through mixed-precision implementations to reduce communication volume and iteration-adaptive redistribution of CI-vector segments based on measured wall times from preceding Davidson iterations to reduce residual load imbalance.

Overall, this work establishes a scalable diagonalization framework for large selected determinant spaces and provides evidence for the structural compactness of the TPB representation.

\section*{Supplementary Material}
The Supplementary Material includes detailed computational timings (Tables S1–S2), the potential energy surfaces of CN (Table S3), and additional coefficient-distribution plots (Figure S1)

\section*{Acknowledgments}
The authors acknowledge Muneaki Kamiya for proposing the subspace restart strategy in the Davidson algorithm, which helped reduce memory usage. 
This work utilized computational resources of supercomputer Fugaku(Project ID: hp250291) and the Hokusai Big Waterfall2 system at RIKEN (Project IDs: RB230090 and RB250029).

\section*{Author Declarations}
\subsection*{Competing interests}
The authors declare that they have no competing interests.

\subsection*{Author contributions}
E.X. conceived the TBSCI framework, formulated the SCI-derived bitstring selection scheme, designed the Hamiltonian application algorithm and the MPI communication strategies, implemented the code, and performed the majority of the calculations. 
W.D. provided technical feedback during the implementation phase. 
H.P. carried out the reference SCI calculations using DICE. 
T.N. initiated and supervised the research project.

E.X., W.D., and H.P. co-wrote the manuscript. 
All authors contributed to validation experiments, discussions, and manuscript revisions.

\subsection*{Data availability}
Input files and representative output data are available from the corresponding author upon a reasonable request.

\section*{Code availability}
Executable binaries of the TBSCI eigensolver used in this work, together with representative input and output data for the reported benchmark calculations, are publicly available via a GitHub repository:
\url{https://github.com/enhua-xu/TBSCI-202603}.

\bibliography{refs.bib}

\end{document}